\documentclass[aps,prb,twocolumn,floatfix,superscriptaddress,showpacs]{revtex4-1}
\usepackage[centertags]{amsmath}
\usepackage{graphicx}
\usepackage{epstopdf}
\usepackage{siunitx}
\usepackage{amsfonts}
\usepackage{amssymb}
\usepackage{bm}
\usepackage{dcolumn}
\usepackage{epstopdf}

\newcommand{\Zgc}{\mathcal{Z}}
\newcommand{\mugc}{\mu_{\mathrm{gc}}}
\newcommand{\eref}[1]{Eq.~(\ref{#1})}
\newcommand{\figref}[1]{Fig.~\ref{#1}}
\renewcommand{\vec}[1]{\boldsymbol{#1}}

\begin{document}

\title{Quantum Monte Carlo measurement of the chemical potential of helium-4}
\author{C. M. Herdman}
\email{Christopher.Herdman@uvm.edu}
\affiliation{Department of Physics, University of Vermont, Burlington, Vermont 05405, USA}
\author{A. Rommal}
\affiliation{Department of Physics, Muhlenberg College, Allentown, Pennsylvania 18104, USA}
\author{A. Del Maestro}
\affiliation{Department of Physics, University of Vermont, Burlington, Vermont 05405, USA}
\date{\today}

\begin{abstract}
A path integral Monte Carlo method based on the worm algorithm has been developed to compute the chemical potential of interacting bosonic quantum fluids.  By applying it to finite-sized systems of helium-4 atoms, we have confirmed that the chemical potential scales inversely with the number of particles to lowest order. The introduction of a simple scaling form allows for the extrapolation of the chemical potential to the thermodynamic limit, where we observe excellent agreement with known experimental results for helium-4 at saturated vapor pressure. We speculate on future applications of the proposed technique, including its use in studies of confined quantum fluids.
\end{abstract}


\maketitle

\section{Introduction}

The chemical potential $\mu$, of a fluid measures the tendency of particles to diffuse as a function of spatial position and sets a characteristic energy scale: that needed to add a single particle to the system at constant temperature.  As an intensive thermodynamic quantity, it is most directly defined as an energy difference:
\begin{equation}
\mu\left(N,T \right) \equiv F\left(N+1,T \right)-F\left(N,T\right), 
\label{mu_eq}
\end{equation}
where $F(N,T)$ is the Helmholtz free energy of a $N$-particle system at temperature $T$.  A spatial gradient in the chemical potential can be established via a pressure ($\Delta P$) or temperature ($\Delta T$) difference as defined by the Gibbs-Duhem relation:
\begin{equation}
\Delta \mu = \frac{V}{N} \Delta P - \frac{S}{N} \Delta T
\label{eq:Gibbs}
\end{equation}
where $V$ is the volume and $S$ the entropy.  Particles will diffuse to the region of low chemical potential and an equilibrium mass current can be established.  In this way, the chemical potential is the analog of the electrical potential for neutral particles.  Unlike other intensive thermodynamic quantities such as temperature or pressure, $\mu$ is often not directly fixed or measured experimentally, and thus a complete understanding of its nature  may provide new insights when fluids are in equilibrium with other phases.  Additionally, $\mu$ determines the locations of first order phase transitions, as well as the nature of mixtures when impurities are present in the fluid.

At low temperature,  even in the absence of interactions, $\mu$ is a sensitive probe of quantum mechanical behavior. For example, an ideal Bose gas at $T=0$ has $\mu=0$, indicating the presence of a zero momentum Bose-Einstein condensate, while for free fermions, $\mu=\varepsilon_\textrm{F}$ showcasing the existence of a Fermi surface where single particle states are filled up to an energy $\varepsilon_\textrm{F}$. 

As is evident from its definition in Eq.~(\ref{mu_eq}), $\mu$ is sensitive to finite-size effects, and it is thus interesting to investigate its value in a mesoscopic regime where interference occurs between length scales set by quantum coherence and the sample geometry.   Such a regime can be experimentally accessed through dimensional confinement of any system displaying macroscopic quantum phenomena.  One such system (and the focus of this paper) is a quantum fluid of bosonic $^4$He, where due to a competition between kinetic and potential energy in the bulk, there is no transition to a solid phase at atmospheric pressure down to absolute zero temperature. Instead, $^4$He undergoes a phase transition to a superfluid at $T_\lambda \simeq \SI{2.172}{\kelvin}$ characterized by macroscopic phase coherence yielding zero viscosity and persistent quantized mass flow in a toroidal geometry \cite{Leggett:2006qf}.  

Recent advances in nanofabrication techniques are opening up new avenues for the study of dimensional crossover in $^4$He and an understanding of the important role played by the chemical potential is emerging.  When confined to two spatial dimensions, relative changes in the chemical potential of $^4$He from its bulk value can be used to determine the thickness of a quasi-two-dimensional film in equilibrium with its vapor\cite{Gasparini2008a}. As the thickness of the film is reduced to only a few atomic layers, unequivocal signatures of the Berezinskii-Kosterlitz-Thouless \cite{Berezinskii:1971uu, Kosterlitz:1972wf, *Kosterlitz:1973uj} vortex unbinding transition can be observed \cite{Gasparini2008a, Agnolet:1989kt, Steele:1993ps}.  A fixed chemical potential difference between two reservoirs of helium-4 connected by quasi-one-dimensional weak links at $T\lesssim T_\lambda$ drives Josephson oscillations with frequency $\omega = \Delta \mu  / \hbar$ \cite{Hoskinson:2005km}. More recently, a pressure difference at constant temperature, leading to a spatial gradient in the chemical potential was used to study $^4$He mass flow through a single nanohole \cite{Savard:2011fe} bringing into question the commonly held definition of a superleak's impenetrability to normal fluid flow. In such extreme confinement, the total number of helium atoms may number in the tens of thousands, and a microscopic understanding of the chemical potential may prove fruitful in interpreting experimental results. 

Monte Carlo methods provide some of the most useful theoretical tools for the study of interacting fluids and readily allow for the scalable computation of expectation values of extensive observables. Intensive thermodynamic quantities that are defined by a free energy difference (like $\mu$) require more sophisticated algorithms, as they are not defined in terms of local observables alone. Extended ensemble Monte Carlo methods such as umbrella sampling can provide numerical access to free energy differences via the calculation of relative probabilities, \cite{Landau2009c,Binder2010a} but they are numerically costly and simpler, more efficient algorithms are desirable. The definition of the chemical potential in \eref{mu_eq} is most directly accessed in the canonical ensemble, where the number of particles is fixed as an extensive parameter and $\mu$ is the conjugate intensive observable. Conversely in the grand canonical ensemble, $\mu$ is a parameter, and $N$ (or the density $n$) becomes an observable. For systems where the grand canonical ensemble is the most appropriate choice for calculations (\emph{e.g.} systems with phase coexistence), it is often necessary to tune the chemical potential to produce a target density. Consequently, having quantitative results for the finite-size scaling of the chemical potential is of great practical importance for numerical simulations.

The worm algorithm (WA) \cite{Boninsegni2006,*Boninsegni2006a} allows for grand canonical path integral Monte Carlo \cite{Ceperley1995} (PIMC) simulations of interacting bosonic quantum fluids. In the WA, the ergodicity in particle number sectors can be treated as an effective \emph{extended ensemble} method for canonical ensemble calculations; in particular the relative probability of different particle number sectors determines the free energy difference between them and therefore the chemical potential. In this paper we present a Monte Carlo method exploiting the worm algorithm to compute the chemical potential of interacting bosonic quantum fluids and apply it to the study of the finite-size scaling of the chemical potential of liquid helium-4.  We begin with an brief introduction to the determination of the chemical potential in classical fluids, before presenting details of how it can be precisely measured in high density quantum liquids via quantum Monte Carlo methods.   Numerical results from a finite-size system of helium-4 atoms demonstrates the expected $1/N$ scaling of $\mu$ with a pre-factor that is related to the compressibility of the quantum fluid. When extrapolated to the thermodynamic limit, the calculated values of $\mu$ as a function of temperature are in excellent agreement with known and inferred experimental results for bulk $^4$He at saturated vapor pressure.

\section{Finite-size scaling of the chemical potential in classical fluids}

For a $d$-dimensional ideal classical fluid confined inside a hypercube of side $L$, the chemical potential is only a function of density and temperature:
\begin{equation}
    \mu_0(n,T) = k_{\text{B}} T \ln \left( n \lambda^{d} \right),
\label{eq:mu0}
\end{equation}
where $n = N/L^d$ is the number density, $k_{\text{B}}$ the Boltzmann constant, and $\lambda$ is the thermal wavelength:
\begin{equation}
    \lambda\equiv \sqrt{\frac{2 \pi \hbar^2}{m k_{\text{B}}T}}.
\end{equation}
Adding interactions generates finite-size scaling of $\mu(N,T)$ at constant density, which Siepmann {\it et al.} computed to leading order with the result~\cite{Smit1989,Siepmann1992}: 
\begin{align}
    \mu \left( N,T \right) &= \mu_0 + \frac{1}{2N} \left( \frac{\partial P}{\partial n} \right) \nonumber \\  
                         & \, \times \left\{ 1 - k_\mathrm{B} T \left[\left( \frac{\partial n}{\partial P}\right) +\left(\frac{\partial^2 P}{\partial n^2}\right) \left( \frac{\partial P}{\partial n }\right)^{-2}\right] \right\}, 
\label{muNclass}
\end{align}
%
demonstrating that the leading deviation from the ideal fluid value scales as $1/N$ where the coefficient is determined by the isothermal compressibility $\kappa = (1/n)(\partial n/\partial P)_T$ as well as $\partial^2 P / \partial n^2$. However, the derivation of \eref{muNclass} may be invalid when the interaction potential has an attractive part as $( \partial P/ \partial n )$ can vanish \cite{Siepmann1992}. More generally, the form of the scaling of $\mu$ and the validity of Eq.~(\ref{muNclass}) can be investigated numerically.

\subsection{Widom particle insertion method for classical fluids}

For Monte Carlo simulations of classical fluids, Widom proposed a method to compute the chemical potential in the canonical ensemble~\cite{Widom1963}. The starting point is the partition function of a $d$-dimensional classical fluid, which can be written as
\begin{equation}
Z_N = \frac{1}{N! \lambda^{Nd}}  Q_N \label{Zcl}
\end{equation}
with $Q_N$ the configuration integral involving only the potential energy $U$:
\begin{equation}
    Q_N = \int \mathcal{D}\, \vec{r} \mathrm{e}^{-\beta U}
\end{equation}
where $\int \mathcal{D}\vec{r} \equiv \prod_{i=1}^{N} \int d^d r_i$ and $\beta=1/k_\mathrm{B}T$ is the inverse temperature.  For an ideal fluid, $U = 0$ so $Q_N = L^{Nd}$ and Eq.~(\ref{eq:mu0}) is immediately recovered. Consequently, in the presence of interactions ($U\ne0$), the excess chemical potential only involves a ratio of $Q_N$ and $Q_{N+1}$:
\begin{equation}
    \mu \left( N, T \right) = \mu_0\left( n,T \right) - k_\mathrm{B} T \log{\frac{1}{L^d} \frac{Q_{N+1}}{Q_N}}.
\end{equation}
Widom showed that the ratio of configuration integrals is related to the expectation value of a canonical observable:
\begin{equation}
    \frac{Q_{N+1}}{Q_N} = L^d \left \langle \left \langle \mathrm{e}^{-\beta \Psi\left(\vec{r}\right)} \right \rangle_{\boldsymbol{r}} \right \rangle_N \label{eq_Widom}
\end{equation}
where $\Psi(\vec{r})$ is the potential energy of an extra particle at spatial position $\vec{r}$ interacting with the other $N$ particles in a canonical ensemble average. The average of the exponential is taken over space as well as the N-particle ensemble. One can thus compute the chemical potential in a canonical Monte Carlo simulation by measuring this ``Widom particle insertion'' observable. Of course this method is limited to classical fluids where the partition function may be factorized as in equation~(\ref{Zcl}) and fails at high densities or when the particles have impenetrable cores and the statistical weight of insertion becomes exponentially small.

This method and its variants has been used to study the finite-size scaling of low-density classical fluids with hard-sphere~\cite{Adams1974}, and Lennard-Jones interactions~\cite{Shing1981,Shing1982,Smit1989a,Attard1997,Heinbuch1987}. We note that previous studies have found a sensitivity of the chemical potential to the cutoff radius used for the long range interaction tail~\cite{Heinbuch1987}. It is natural to ask if these algorithms can be extended beyond the classical domain for application to quantum fluids.  Here, Monte Carlo configurations of atoms are extended to \emph{worldlines} and the analog of the Widom method would require inserting a non-local object.  Such an insertion would necessarily have an exponentially small weight with the algorithm becoming intolerably inefficient at high densities.  However, the worm algorithm (discussed below) allows for the efficient insertion and removal of particles via only local updates and thus offers a novel platform for computing the chemical potential in quantum fluids.

\section{Computing the chemical potential with Quantum Monte Carlo}

The worm algorithm~\cite{Boninsegni2006,*Boninsegni2006a} is a modern variant of path integral Monte Carlo \cite{Ceperley1995} that allows for efficient grand canonical simulations of bosonic quantum fluids.  Any extensive observable that can be written in the position basis (\textit{e.g.} number of particles) can be easily calculated, while the conjugate intensive quantities (\textit{e.g.} the chemical potential) are parameters of the simulation. This is because extensive quantities can be computed directly from the expectation value of the observable, whereas intensive quantities are generally defined by derivatives of the free energy. While the free energy can be computed with thermodynamic integration, such a procedure is very computationally expensive and generates large statistical errors. It is thus desirable to search for other methods that allow for the accurate computation of intensive quantities such as the chemical potential.

\subsection{Worm algorithm path integral Monte Carlo}

Path integral Monte Carlo methods \cite{Ceperley1995} use a configuration space of particle imaginary-time worldlines in $d+1$ dimensions (for a system of $d$ spatial dimensions) to statistically sample the many-body density matrix of any interacting system in the spatial continuum that can be described by a Hamiltonian of the general form:
\begin{equation}
    H = \sum_{i=1}^N \left({ - \frac{\hbar^2}{2m_i} \nabla_i^2 + V_i}\right) + \sum_{i<j} U_{ij},
\label{eq:ham}
\end{equation}
where $m_i$ is the mass of a particle located at position $\vec{r}_i$, $V_i$ is an external potential and $U_{ij}$ is any two-body interaction.  In canonical PIMC, configurations of closed worldlines are sampled from the Boltzmann distribution, $\mathrm{e}^{-\beta {H}}$, which allows for the calculation of canonical expectation values. In the worm algorithm, configurations with both open and closed worldlines are allowed~\cite{Boninsegni2006,*Boninsegni2006a} and such open worldlines, or \emph{worms}, sample off-diagonal elements of the density matrix. When an open worldline winds around the imaginary-time axis and closes, this will return the system to a diagonal worldline configuration with an additional particle: $N \rightarrow N+1$. Conversely, it may be energetically favorable to open a worldline, creating a worm, which can shrink until it is completely removed from the configuration with $N \rightarrow N-1$.  The WA therefore naturally operates in the grand canonical ensemble, as the presence of worms leads to fluctuations in the total particle number. In addition to allowing for grand canonical simulations using only local updates, the worm algorithm yields efficient permutation sampling, as well as access to the imaginary-time Green function. For our purposes, we are most interested in the WA as an efficient grand canonical PIMC method to study interacting bosonic quantum fluids. 

\subsection{Computing the chemical potential in Monte Carlo}

To directly compute the intensive chemical potential, we need access to the free energy difference defined in Eq.~(\ref{mu_eq}):
\begin{align}
    \mu(N,T) &= F(N+1,T) - F(N,T) \nonumber \\
        &= - k_\mathrm{B} T \log{\frac{Z_{N+1}}{Z_N}},
\label{eq:muZ}
\end{align}
where $Z_N = \mathrm{Tr}\, \mathrm{e}^{-\beta H}$ is the canonical partition function for the $N$-particle system. As previously mentioned, one may compute such free energy differences between two states labeled $a$ and $b$ via thermodynamic integration~\cite{Landau2009c,Binder2010a} using a relation of the form:
\begin{equation}
\Delta F = F_b - F_a = \int_{\eta_a}^{\eta_b} d \eta \left( \frac{\partial F}{\partial \eta} \right)
\end{equation}
where $\eta$ is a parameter. If the expectation value of $\partial F / \partial \eta$ is computed for a discrete set of $\eta$ interpolating between $\eta_a$ and $\eta_b$ then $\Delta F$ may be estimated by numerical integration. Such methods have been used for the chemical potential, where $\eta$ is taken to be a coupling strength between particles where $\eta=0$ is the free particle limit~\cite{Kristof2007}. In general, thermodynamic integration is quite computationally expensive as sufficient $\eta$ points must be chosen to minimize the systematic error in the numerical integration as well as the fact that the statistical errors accumulate in the integration.

To avoid performing such a numerically costly procedure, we can instead treat the grand canonical ensemble of the WA as an extended ensemble for the particle number sectors that are accessed through thermal and quantum fluctuations.  In this way, the WA can be used to directly measure the free energy difference between two canonical ensembles at the same temperature, and thus compute the chemical potential in a single grand canonical calculation.

For a system whose Hamiltonian, ${H}$ conserves the number of particles, we can decompose the grand canonical partition function, $\Zgc$, in terms of canonical partition functions $Z_N$:
\begin{align}
    \mathcal{Z} &= \mathrm{Tr}\, \mathrm{e}^{-\beta \left({H} - \mugc N \right)} \nonumber \\
                &= \sum_{N=0}^{\infty} \mathrm{e}^{\beta \mugc N} Z_N,
\end{align} 
where $\mugc$ is the chemical potential of the grand canonical ensemble. The ratio of the canonical to the grand canonical partition function, is an observable which is directly measurable in a Monte Carlo calculation,
\begin{equation}
    \frac{Z_N}{\Zgc }=  \mathrm{e}^{-\beta \mugc N}\left \langle \delta_N \right \rangle_\Zgc. \label{ZoverZgc}
\end{equation}
The expectation value $\langle \delta_N \rangle_\Zgc$ is the probability of the grand canonical simulation having $N$ particles, $P\left(N\right)$, which is readily computed by tabulating a histogram of the values of $N$. Therefore, in a grand canonical simulation, the ratio of the partition functions of different particle number sectors can be computed from $P\left(N\right)$. Consequently, the chemical potential can be directly measured from $P\left(N\right)$:
\begin{equation}
    \mu \left( N, T \right) = \mugc - k_\mathrm{B} T \log  \frac{P \left(N+1\right)}{P \left( N \right) } 
\label{muWAPIMC}
\end{equation}
In \eref{muWAPIMC}, $\mugc$ is a parameter of the grand canonical simulation that can be chosen such that the density of interest is efficiently sampled. However, the physical canonical chemical potential $\mu(N,T)$ is independent of $\mugc$, so long as the number sector $N$ is efficiently sampled by the grand canonical calculation.  If a single fixed density $n$ is of interest, only the ratio $P(N+1)/P(N)$ is required to determine $\mu(N,T)$. Therefore we may improve the efficiency of computing $\mu(N,T)$ by limiting the particle number fluctuations to $N\pm1$. This can be easily implemented in a WA simulation by rejecting any Monte Carlo updates which increase or decrease the particle number by more than one from the target number of particles $N = n L^d$. This changes the values of $P(N)$ that are computed in the simulation, but detailed balance ensures that the ratio $P(N+1)/P(N)$ remains unaffected by this restricted sampling.

\section{Results: finite-size scaling of the chemical potential of helium-4}

Although the methods proposed here can be directly applied to any many-body system described by Eq.~(\ref{eq:ham}), we have have chosen to exhibit their efficacy in computing the finite-size scaling of the chemical potential for liquid helium-4 at saturated vapor pressure (SVP). This will allow for direct benchmarking and comparison with experimental results at low temperature \cite{Niemela1995, Pettersen1995}.  We use the Aziz potential~\cite{Aziz1979} to model the inter-atomic interactions $U_{ij}$ in liquid helium-4 with $\hbar^2/(2 m k_\mathrm{B}) \simeq \SI{6.055}{\angstrom^2\kelvin}$, set the external potential to zero $(V_i=0)$, and work in a $d=3$ cubic simulation cell of side $L$ with periodic boundary conditions. In the remainder of this paper, we will measure all energies in kelvin and thus set $k_\mathrm{B}=1$.

\begin{figure}[t]
\begin{center}
\includegraphics[width=\columnwidth]{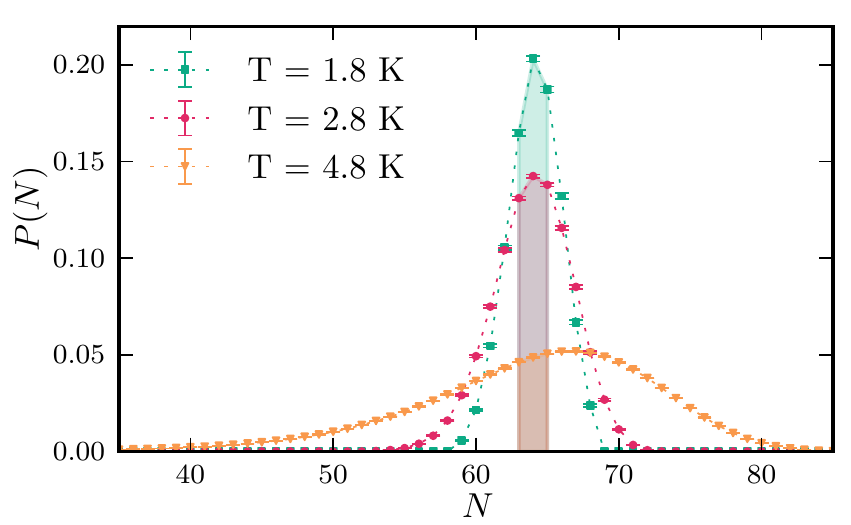}
\caption{\label{fig:PofN}(color online). The particle number probability distribution $P(N)$ computed with quantum Monte Carlo for $^4$He at $T=$\SIlist{1.8;2.8;4.8}{\kelvin} at saturated vapor pressure with $L=$\SIlist{14.304;14.387;15.702}{\angstrom} and $\mugc=$\SIlist{-4.3;-5.1;-8.2}{\kelvin} such that $N_\mathrm{SVP}(T) \approx 64$. The shaded area indicates the restricted region of particle numbers that were sampled.
}
\end{center}
\end{figure}

The canonical chemical potential in Eq.~(\ref{muWAPIMC}) can be directly evaluated from the relative probability that a WA simulation at temperature $T$ has $N$ or $N+1$ particles at fixed volume.  We have performed such simulations for a range of temperatures both above and below $T_\lambda$ with a few representative examples of the full particle number probability distribution function shown in \figref{fig:PofN}.  
As described above, the grand canonical chemical potential $\mugc$ enters as a parameter that can be tuned to efficiently sample observables at the desired density (in this case, SVP).  This corresponds to ensuring that the maximum of $P(N)$ in \figref{fig:PofN} occurs near $N_\mathrm{SVP}(T)$ such that $n_{\mathrm{SVP}}(T) = N_\mathrm{SVP}(T)/L^3$; however, no fine tuning is required, as it is only the efficiency of the sampling of $P(N+1)/P(N)$ that requires this ratio to be of order one. Additionally, near a first order phase transition, where phase coexistence leads to a double peak structure of $P(N)$, it can be beneficial to tune $\mugc$ away from the transition to avoid the large fluctuations in $N$. In practice, we restrict our simulations to number fluctuations of $\pm 1$ indicated by the shaded regions in Fig.~\ref{fig:PofN}.

\begin{figure}[t]
\begin{center}
\includegraphics[width=\columnwidth]{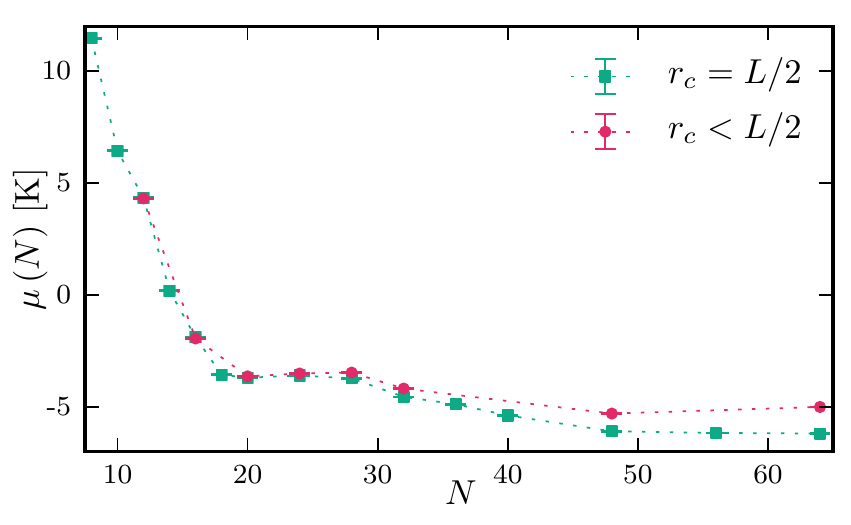}
\caption{\label{fig:muvNCO}(color online). An interaction potential cutoff length $r_c < L/2$ causes the chemical potential to prematurely saturate as a function of the number of particles $N$. Simulations were performed for $^4$He atoms at $T=\SI{2.8}{\kelvin}$ inside a cube with periodic boundary conditions at $n_{\mathrm{SVP}} = \SI{0.0214922}{\angstrom^{-3}}$. The data for $r_c < L/2$ corresponds to $r_c = \SI{7}{\angstrom}$ which is less than $L/2$ for the larger system sizes shown.}
\end{center}
\end{figure}

In many PIMC simulations employing an Aziz-like potential, the $N^2$ scaling of $U_{ij}$ in Eq.~(\ref{eq:ham}) is reduced by choosing a hard cutoff length $r_c$ for the Van der Waals tail with the interactions being neglected beyond this distance. The use of a cutoff, in combination with a spatial lookup table \cite{Allen:1989cs} can drastically improve simulation efficiency. However, we find that the chemical potential, calculated via \eref{muWAPIMC} is highly sensitive to the cutoff and plateaus to a value that is substantially above the expected bulk value when  $r_c < L/2$. This effect can be seen in \figref{fig:muvNCO}, where the chemical potential is plotted as a function of the total number of particles for a system with  $n_{\mathrm{SVP}}(T=\SI{2.8}{\kelvin}) = \SI{0.0214922}{\angstrom^{-3}}$ for two values of the potential cutoff $r_c = \SI{7}{\angstrom} < L/2$ and $r_c = L/2$. 
Not surprisingly, the energy required to add a particle to the system is sensitive to the details of the potential tail, and a premature saturation occurs, indicating a breakdown of finite-size scaling.  As our goal it to accurately measure the finite-size scaling of the chemical potential, we have chosen to use the full long distance tail of the Aziz potential in our simulations and are thus limited to systems composed of less than $150$ particles.  The unavoidable Trotter error is constrained to be smaller than statistical uncertainties through the use of a short-time imaginary-time propagator that is accurate to fourth order in the imaginary time step $\Delta \tau$~\cite{Jang2001} which we set as $\Delta \tau = \SI{0.004}{\kelvin^{-1}}$. 

\begin{figure}[t]
    \begin{center}
    \includegraphics[width=\columnwidth]{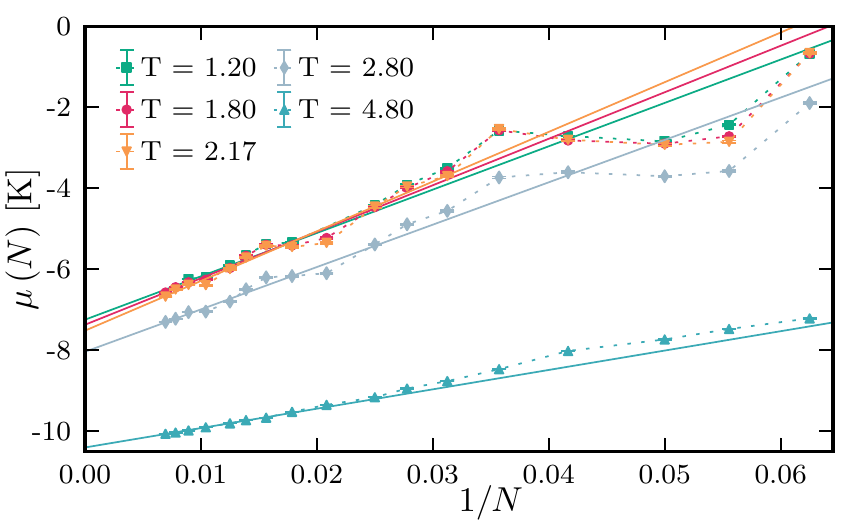}
    \caption{(color online). The constant density finite-size scaling of the chemical potential of $^4$He at saturated vapor pressure for various temperatures via large scale quantum Monte Carlo simulations. The solid line represents the best $1/N$ fit to the linear region of the data. }
    \label{fig:1onN}
    \end{center}
\end{figure}

Fixing the cutoff at $r_c = L/2$, we have computed the dependence of the chemical potential on the number of particles $\mu(T,N)$ for temperatures above and below $T_\lambda$ with results shown in Fig.~\ref{fig:1onN}. 
The finite-size scaling was done at constant density, chosen to be the experimentally determined thermodynamic density at saturated vapor pressure at each temperature (see Table~\ref{tab:data}). For larger system sizes, there is a clear $1/N$ scaling, as expected from the theory of classical fluids. Additionally, we see oscillations about this $1/N$ scaling that decrease in amplitude as the system size increases, but are still noticeable at the largest system sizes $N=144$. The presence of such non-monotonic scaling with system size is due to the interplay between the discrete nature of the particle number and confinement of the fluid inside a cubic box that explicitly breaks the rotational symmetry of the Aziz interaction potential $U_{ij}$.

\begin{figure}
\begin{center}
\includegraphics[width=\columnwidth]{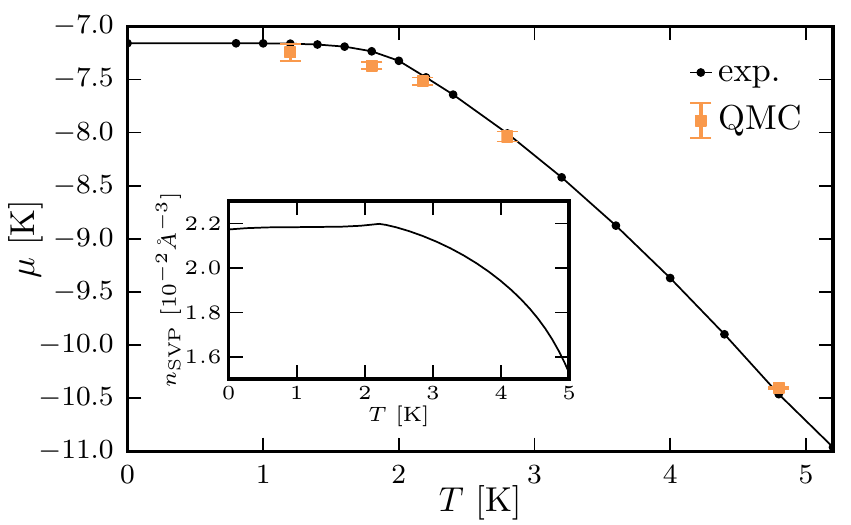}
\caption{\label{fig:mu0vsExpt} (color online). The chemical potential $\mu$ in the thermodynamic limit of $^4$He at low temperature from an extrapolation of quantum Monte Carlo data using the finite-size scaling form of \eref{eq:fssmu}. The reported errorbars include both statistical and fitting uncertainties. Experimentally determined values for bulk $^4$He from Refs.~[\onlinecite{Niemela1995}] and [\onlinecite{Pettersen1995}] are included as filled circular symbols for comparison and show agreement withing a few percent. {\it Inset}: Experimental number density at saturated vapor pressure of $^4$He taken from Ref.~[\onlinecite{Niemela1995}]}
\end{center}
\end{figure}
To test the accuracy of our algorithm and make contact with experimental results for the bulk SVP values of the chemical potential of helium-4, we can exploit the $1/N$ finite-size scaling seen in Fig.~\ref{fig:1onN} to extrapolate to the thermodynamic limit. Assuming the scaling form:
\begin{equation}
\mu\left(N,T\right) = \mu \left(T \right) + \frac{c\left(T\right)}{N}
\label{eq:fssmu}
\end{equation}
we have performed a linear regression of $\mu(N,T)$ for large $N$. Fig.~\ref{fig:mu0vsExpt} shows the extrapolated values of $\mu(T) \equiv \lim_{N\to\infty} \mu (N,T)$ along with experimental results from  Refs.~[\onlinecite{Niemela1995}] and [\onlinecite{Pettersen1995}] with the numerical values given in Table~\ref{tab:data}.
\begin{table}[t]
\begin{center}
    \renewcommand{\arraystretch}{1.2}
  \begin{tabular}{ | c | c | c | c | c |}
    \hline
    $T~[K]$	&$n~[\si{\angstrom^{-3}}]$& $\mu~[K]$		& $\mu^\mathrm{exp}~[K]$ 	& $c~[K]$ 	\\ \hline
    $1.20$ 	& $0.021833$	&  $-7.25\pm0.08$		& $-7.1638$ 			& $107\pm8$		\\ \hline 
    $1.80$ 	& $0.021869$	& $-7.37\pm0.03$ 		& $-7.2356$			& $115\pm4$		\\ \hline
    $2.18$ 	&  $0.021983$	& $-7.52\pm0.03$		& $-7.4655$ 	& $123\pm4$		\\ \hline
    $2.80$ 	&  $0.021492$	& $-8.04\pm0.05$ 		& $-8.0083$ 			& $105\pm6$		\\ \hline
    $4.80$	&  $0.016531$	& $-10.400\pm0.009$	& $-10.4607$ 			& $47.8\pm0.9$	\\ \hline
  \end{tabular}
\end{center}
\caption{\label{tab:data} Numerical values for estimates of the bulk chemical potential $\mu$ and scaling pre-factor $c$, of $^4$He obtained from a $1/N$ fit to the linear regime of the quantum Monte Carlo data displayed in Fig.~\ref{fig:1onN}. The number densities $n$, and experimental values of the chemical potential in the thermodynamic limit $\mu^{exp}$ are taken from Refs.~[\onlinecite{Niemela1995}] and [\onlinecite{Pettersen1995}]. The experimental value of the chemical potential for $T=\SI{2.18}{\kelvin}$ was determined by linear interpolation of the data. }
\end{table}
The precision of this extrapolation is limited by the system sizes studied and the oscillations about the $1/N$ scaling due to the residual finite-size density correlations. Despite this, we see agreement between our numerically determined values of $\mu(T)$ and the experimental values to within a few percent.  

\section{Discussion}

In this paper, we have demonstrated the use of worm algorithm path integral Monte Carlo as an effective extended ensemble method, able to efficiently compute the finite-size value of the intensive canonical chemical potential.   We have found an excellent correspondence between simulations consisting of only a few hundred atoms extrapolated to the thermodynamic limit, and bulk experiments on helium-4, demonstrating the feasibility of using this approach to determine the chemical potential in real, experimentally accessible quantum fluids and gases. 

It is important to point out that the inventors of the worm algorithm have already brought attention to, and demonstrated its ability to determine the chemical potential through knowledge of the equilibrium density as a function of the grand canonical parameter $\mugc$: $n(\mugc)$ \cite{Boninsegni:2006gca}. Interpolation can then be performed by independently measuring the isothermal compressibility $\kappa$, determined from number fluctuations: $n^2 \kappa = d n / d\mu = \langle \left( N - \langle N^2 \rangle \right) \rangle \beta L^{-d}$.  With this information, Boninsegni~\emph{et al.} have used the known freezing density of solid helium-4, $n_\mathrm{freeze} = \SI{0.02599}{\angstrom^{-3}}$ and find the chemical potential at $T=\SI{0.25}{\kelvin}$ to be $\mu_\mathrm{freeze} = 0.06 \pm \SI{0.04}{\kelvin}$. The method we present here offers an alternative approach to such calculations, that may prove to be a more efficient in some circumstances. In particular, our approach only requires sampling $P(N_0)$ and $P(N_0+1)$, where $N_0$ corresponds to the target number of particles, whereas computing the compressibility as in Ref.~[\onlinecite{Boninsegni2006}] requires sampling the full $P(N)$ for all $N$ for several values of $\mugc$ to perform the interpolation.

Having access to an efficient method for computing the chemical potential in quantum Monte Carlo will allow for its determination in experimental systems where it is not directly measurable, and a host of future applications of this technique are apparent.  For example, it may now be practical to accurately locate lines of phase coexistence in quantum fluids and ultra-cold gases by numerical simulations. Such information may be especially useful in mixtures of helium-3 and helium-4 as well as multi-component Bose gases.  For low-dimensional helium-4, confined to flow through hollow channels inside mesoporous silica \cite{Ikegami:2003cb}, the exact value of the chemical potential is intricately linked to the thickness of wetting layers in the substrate.  A quantitative understanding of these layers is essential in the interpretation of subsequent measurements of quasi-$1d$ dynamic superfluidity \cite{Taniguchi:2013us}. 

Exact knowledge of the finite-size scaling of the chemical potential may also provide a new tool to analyze the properties of theoretical models of interacting bosons in the one-dimensional continuum. Much is already understood about such models \cite{Cazalilla:2011dm}, including their universal description at low energies and long wavelengths in terms of the emergent quantum hydrodynamics known as Luttinger liquid theory.  The resulting effective Hamiltonian has a single parameter $K$, which describes a crossover between a superfluid and mass density wave lacking any long range order, with all correlation functions decaying algebraically.  The value of $K$, which is a function of the chemical potential (or density),  can be determined exactly for some simple models of bosons, including those with hard-core \cite{Girardeau:1960ff} or delta-function \cite{Lieb:1963ik} interactions. For more realistic systems, with potentially long range dipole interactions, it can be computed by comparing the results of quantum Monte Carlo simulations with the predictions of Luttinger liquid theory \cite{DelMaestro:2011ll}.  These methods rely on fitting to the complete $L$ and $T$ scaling form of the number probability distribution $P(N)$ and are thus very sensitive to anharmonic finite-size corrections to the quadratic Luttinger liquid Hamiltonian \cite{DelMaestro:2010bb}.  By instead computing $\mu(N,T)$ via the ratio of $P(N+1)/P(N)$ in Eq.~(\ref{muWAPIMC}), these corrections will drop out to lowest order, providing a considerably more accurate and robust route to the determination of $K$ for one-dimensional bosons via quantum Monte Carlo.

In conclusion, the worm algorithm can be directly exploited to measure intensive thermodynamic quantities, such as the chemical potential, without the need for a cumbersome and potentially error-prone thermodynamic integration of numerical simulation data. Moreover, it continues to provide new insights into the microscopic origin of cooperative macroscopic quantum phenomena in the spatial continuum.  

\acknowledgements
We would like to thank M. Boninsegni for discussions related to the use of a potential tail cutoff in path integral simulations.  This research has been supported by the University of Vermont and enabled by the use of computational resources provided by the Vermont Advanced Computing Core supported by NASA (NNX-08AO96G). A.R. acknowledges support for a REU program from the National Science Foundation (No. DMR-1062966) and the hospitality of the University of Vermont where this work was completed.

\bibliographystyle{apsrev4-1}
\bibliography{refs}

\end{document}